\documentclass[twocolumn]{aastex63}
\usepackage[utf8]{inputenc}
\usepackage[a4paper, total={6in, 8in}]{geometry}

\usepackage{amsmath}
\usepackage{graphicx}
\usepackage{esvect}

\def\degree{$^\circ$}
\def\deg{$^\circ$}

\shorttitle{The Vanishing Transits of KOI 120.01}
\shortauthors{Judkovsky et al.}

\begin{document}
\received{March 4, 2020}
\revised{Aug 18th, 2020}


\title{Light-Curve Evolution due to Secular Dynamics \\ and the Vanishing Transits of KOI 120.01}

\correspondingauthor{Yair Judkovsky}
\author[0000-0003-2295-8183]{Yair Judkovsky}
\affiliation{Weizmann Institute of Science, Rehovot, 76100 Israel}
\email{yair.judkovsky@weizmann.ac.il}

\author[0000-0002-9152-5042]{Aviv Ofir}
\affiliation{Weizmann Institute of Science, Rehovot, 76100 Israel}

\author[0000-0001-9930-2495]{Oded Aharonson}
\affiliation{Weizmann Institute of Science, Rehovot, 76100 Israel}
\affiliation{Planetary Science Institute, Tucson, AZ, 85719-2395 USA }


\graphicspath{{./}{figures/}}


\begin{abstract}
Non-Keplerian dynamics of planetary orbits manifest in the transit light-curve as variations of different types. In addition to Transit Timing Variations (TTV's), the shape of the transits contains additional information on variations in the geometry of the orbit. This study presents an analytic approach to light-curve fitting: dynamical variations in the orbital elements are transformed to a light-curve using an analytic function with a restricted set of fitting parameters. Our method requires no N-body integration, resulting in a smaller number of degrees of freedom and a faster calculation. The approach described here is for the case of secular perturbations.
By assuming that the orbital motion is dominated by nodal and apsidal precessions, analytic expressions for the light-curve transit parameters are derived as a function of the orbital variations. 
Detecting and characterizing such dynamical scenarios provides information regarding the possible existence of non-transiting companions, or the non-spherical mass distribution of the host star. The variations may imply forces out of the orbital plane, and thus probe mutual inclinations among components of the system. The derived models successfully reproduce the vanishing transit signals of KOI 120.01, and suggest a possible interesting scenario of a planet orbiting one member of a close-in binary system undergoing unusually rapid nodal regression. The model parameters are degenerate, so we provide relevant information for followup observations, which are suggested in order to place further constraints on this unique \textit{Kepler} object.
\end{abstract}


\section{Introduction}
The \textit{Kepler} mission \citep{BoruckiEtAl2010} has led to a revolution in our knowledge of planetary systems \citep[e.g.][]{BatalhaEtAl2013}), providing information on the orbital architectures and the multiplicity distribution of planetary systems. In particular, it was found that a single multiplicity distribution is insufficient to explain the excess of singly transiting planets systems \citep{BallardJohnson2016,Lissauer2011}, and a second population, consisting of either single, or more inclined planets,  may be required to explain the observations (also known as "the \textit{Kepler} dichotomy"). A recent statistical study of planetary population found it unlikely that this dichotomy is explained solely by a population of real single-planet systems \citep{Matthias2019}.
Statistical analysis of the distribution of transit durations has shown that multiply-transiting systems are consistent with having a roughly co-planar structure \citep{FabryckyEtAl2014}. With accurate determination of stellar densities \citet{XieEtAl2016} used transit duration distribution to fit for distribution of eccentricities (and mutual inclinations, for multiple-transiting systems), and inferred that there is a dichotomy within the singly-transiting planets population, which consists of dynamically hotter and dynamically colder systems.

Non-transiting companions may hide in systems with singly transiting planets particularly in mutually inclined systems \citep{LaiPu2017}, and the former may be probed by their influence on the latter \citep{MillsEtAl2019}.
Secular effects applied by an exterior companion on an inner transiting planet would yield transit variations, usually described as TTV's \citep{AgolSteffenSariClarkson2005, HolmanMurray2005} and Transit Duration Variation (TDV's) \citep{MiraldaEscude2002}. 
Secular precession rates due to a planetary perturber were estimated for many systems \citep[e.g.,][]{MoroMartinEtAl2007,LeePeale2003}. The time scale of the precession period of a planet due to a secular external perturber is $\tau=4P_1/(\mu_2\alpha^2 b_{3/2}^{(1)}(\alpha))$, where $P$ is the orbital period, $\mu_2$ is the perturber/host mass ratio, $\alpha=a_1/a_2$, where $a$ is the semi-major axes of objects, 
and 1 and 2 correspond to the inner and outer planets, respectively, and $b_{3/2}^{(1)}$ is a Laplace coefficient. 
Derivations can be found in \citet[][chapter 7]{SSD1999}, and expressions for nodal precession rates to second order in inclination were introduced by \citet{XuFabrycky2019}. For comparison, Earth's orbital precession period due to only Jupiter is of the order $10^5$ years. For a close-in planet with an 8~day orbit perturbed by an external Jupiter-mass planet with a 35-day orbit, the precession period is of order 500 years.  During \textit{Kepler}'s primary mission time of $\sim$4~years, this corresponds to an accumulated precession angle of $\sim$3\deg, which may be observable in the current data.

Other sources for secular precession are stellar quadruple moment ({\it e.g.,} KOI 13.01, a hot Jupiter which exhibits nodal precession of $\sim$0.7 \degree $\rm yr^{-1}$ \citep{SzaboEtAl2012}), general relativity ({\it e.g,} HD 80606 b \citep{BlanchetEtAl2019}, and Mercury \citep{Einstein1916}),  and at specific configurations, even tidal interactions \citep{WuGoldreich2002}. In these examples, different drivers yield a similar dynamical phenomenon of secular precession.

In this work, an analytic method for photo-dynamical fitting of the light-curve is presented. The light-curve calculation relies on the assumption that the orbital motion is dominated by secular precession, a likely condition for systems in which there are no near-resonant interactions. As described above, multiple physical processes result in similar behaviour (secular precession) and depending on the physical process the result of such a fit may inform us on  the system's orbital elements (e.g. planet-planet interaction) or other properties (e.g. stellar $J_2$). 
There are a few advantages for this method. First, it does not require a massive number of n-body integration iterations. Second, it utilizes only the transiting planet information to construct a model for the orbital motion. Hence, the mass and orbital elements of the perturbing object are not required (which are often unknown since long-period planets rarely transit). Only solutions for the orbital elements of the transiting, perturbed, planet are sought. Third, it establishes a direct link between the parameters of the model and the dynamics of the system with respect to its invariable plane.

Searching for transiting candidates which could be affected by slow, secular, transit variations, we came across the interesting case of KOI 120.01. This object exhibits unusual vanishing transiting signals, which can be explained by this model.

In \S\ref{Transit Parametrization} the mathematical description for transit variations in the presence of secular precession is presented. In \S\ref{KOI 120.01} the method is applied to the vanishing transit signal of KOI~120.01, and a few possible models for explaining the data are presented. The results are summarized and discussed in \S\ref{Discussion}.

\section{Transit Parametrization} \label{Transit Parametrization}
The aim of the following mathematical derivation is to relate the instantaneous orbital elements with respect to the system's invariable plane (inclination $I$, longitude of the ascending node $\Omega$, and argument of periapse $\omega$) to their counterparts measured with respect to the sky plane ($I_{\rm sky}, \omega_{\rm sky}$). The longitude of ascending node with respect to the sky plane ($\Omega_{\rm sky}$) is irrelevant for the transit shape, as long as the stellar disk is radially symmetric. The coordinates of the planet parameterized by the true anomaly $f$, are given by \citep[][page 51]{SSD1999}:

\begin{equation}
    \begin{bmatrix}
    x       \\
    y       \\
    z       
\end{bmatrix}
=
r\begin{bmatrix}
    \cos{\Omega}\cos{(\omega+f)}-\sin{\Omega}\sin{(\omega+f)}\cos{I}       \\
    \cos{\Omega}\cos{(\omega+f)}+\cos{\Omega}\sin{(\omega+f)}\cos{I}       \\
    \sin{(\omega+f)}\sin{I}       
\end{bmatrix},
\end{equation}
where $x,y,z$ are the coordinates of the planet with the convention that $x,y$ is the system's invariable plane, $x$  lies in the sky plane, and $r$ is the distance from the central mass.
The equation describing the sky plane is given by $y\cos{\beta}+z\sin{\beta}=0$, where $\beta$ is the angle between the observer's line of sight and its projection on the invariable plane ($\beta=0$ corresponds to viewing the invariable plane edge-on, such that the sky plane is simply $x-z$). Solving for the intersection of the orbit with the sky plane yields two solutions for the true anomaly $f$ (representing the ascending and descending nodes through the sky plane). This allows for expressing the orbital elements with respect to the sky plane in terms of those with respect to the invariable plane. The sky plane elements are not required for constructing the light-curve, but give a geometrical interpretation revealing the changes in impact parameter and mid-transit velocity. The sky plane elements are

\begin{equation}
    \cos{I_{\rm sky}} = -\sin{I}\cos{\Omega}\cos{\beta}+\cos{I}\sin{\beta},
\end{equation}
\begin{equation}
    \tan{(\omega_{\rm sky}-\omega)}=\frac{\sin{\Omega}\cos{\beta}}{\cos{\Omega}\cos{I}\cos{\beta}+\sin{I}\sin{\beta}}.
\end{equation}

Gradual changes of the angles $\Omega$ and $\omega$ due to nodal regression or apsidal advance will affect the transit shape in a several ways. The varying separation $s$ between the planet and star will affect the impact parameter and hence the depth and duration of the transit and the shape of ingress and egress. The different longitude observed at each transit will yield a different mid-transit tangential velocity $v_{\rm mid}$, changing in turn the transit duration. Lastly, the different time spent by the planet at each longitude will result in Transit Timing Variations (TTV's).

Making use of the relation $f_{\rm mid} = \pi/2-\omega_{\rm sky}$ (with $f_{\rm mid}$ the true anomaly at mid-transit) the transit parameters can be expressed as

\begin{equation}
    s_{\rm mid}=\frac{a(1-e^2)}{1+e\cos{f_{\rm mid}}}=\frac{a(1-e^2)}{1+e\sin{w_{\rm sky}}},
\end{equation}
\begin{equation}
    b=s_{\rm mid}\cos{I_{\rm sky}}
    \label{eq_b}
\end{equation}
and
\begin{equation}
    v_{\rm mid}=na\frac{1+e\sin{\omega_{\rm sky}}}{\sqrt{1-e^2}},
    \label{eq_v}
\end{equation}
where $n=2\pi/P$ is the mean motion, $a$ the semi-major axis and $e$ is the orbital eccentricity. 
The orbital elements $a,e,\Omega,\omega,I,M$ (where $M=n(t-t_p)$ is the mean anomaly and $t_p$ is the time of periapse) are translated instantaneously to invariable plane coordinates ($x,y,z$) through solution of Kepler's equation. Using the angle $\beta$, they are then translated to sky plane coordinates ($X,Y$). From these coordinates and the ratio between transiting planet and host star radii $R_p/R_*$, relative flux is computed, and a model light-curve is constructed \citep{MandelAgol2002}. Instead of using $t_p$, here the parametrization is for the first time of mid-transit $T_{\rm mid}$, which is known to a good accuracy. 

To briefly summarize: given a full set of orbital parameters (and $R_p/R_*$) a model flux point can be directly calculated. Importantly, these orbital elements need not be constant in time, and for a given time evolution of these elements (\textit{i.e.} a list of such sets) a full light curve can be calculated. Below we assume a secular model in which $\Omega, \omega$ are linear in time. Specifically, by assuming all the dynamics is captured by some long term variability in $\Omega, \omega$ a full light-curve can be generated through the parameters $P, T_{\rm mid}, R_p/R_*, a/R_*, e, \omega_0, \dot{\omega}, I, \Omega_0, \dot{\Omega}, \beta$ (where the subscripts "0" refers to the value at $T_{\rm mid}$) without a full N-body integration.

Fitting for the parameters listed above (or any combination of them) can reveal the precession rate of the transiting planet, imprinted in small variations in the photometry of the transits. As the ratio $\dot{\omega}/\dot{\Omega}\sim-2$ (or, equivalently, $\dot{\varpi}/\dot{\Omega}\sim-1$, where $\varpi=\Omega+\omega$ is the longitude of periapse), instead of fitting for $\dot{\omega}$, the fit is performed on the ratio $\dot{\omega}/\dot{\Omega}$. The exact ratio between the secular precession rates depends on the Laplace coefficients, and derivations can be found in \citet[][chapter 7]{SSD1999}. Instead of using the angle $\beta$ which is unconstrained, the angle $I_{\rm sky0}$ was used ({\it i.e.}, the inclination with respect to the sky plane at first mid-transit time). This angle is constrained to be up to a few degrees away from 90$^\circ$ by the existence of observable transits.

\section{Application to KOI 120.01} \label{KOI 120.01}

\subsection{General Description}
\label{subsec:GeneralDescription}

KOI 120.01 (KIC 11869052) is a \textit{Kepler} object of interest exhibiting an unusual pattern of gradually vanishing transit signals. The transit depth decreases over a time span of about 2 years from 2$\cdot10^{-3}$ to 0. Figure \ref{fig:KOI120RawLC} Shows the detrended photometric light-curve as well as a few specific transits and one of the fitted models for visualization (see \S\ref{LC_Fitting}, the different models prodcue very similar light-curves). 

Other than \textit{Kepler}'s data, the following observations of KOI-120 were found in the literature:

\begin{figure*}[h]
    {\includegraphics[width=1\linewidth]{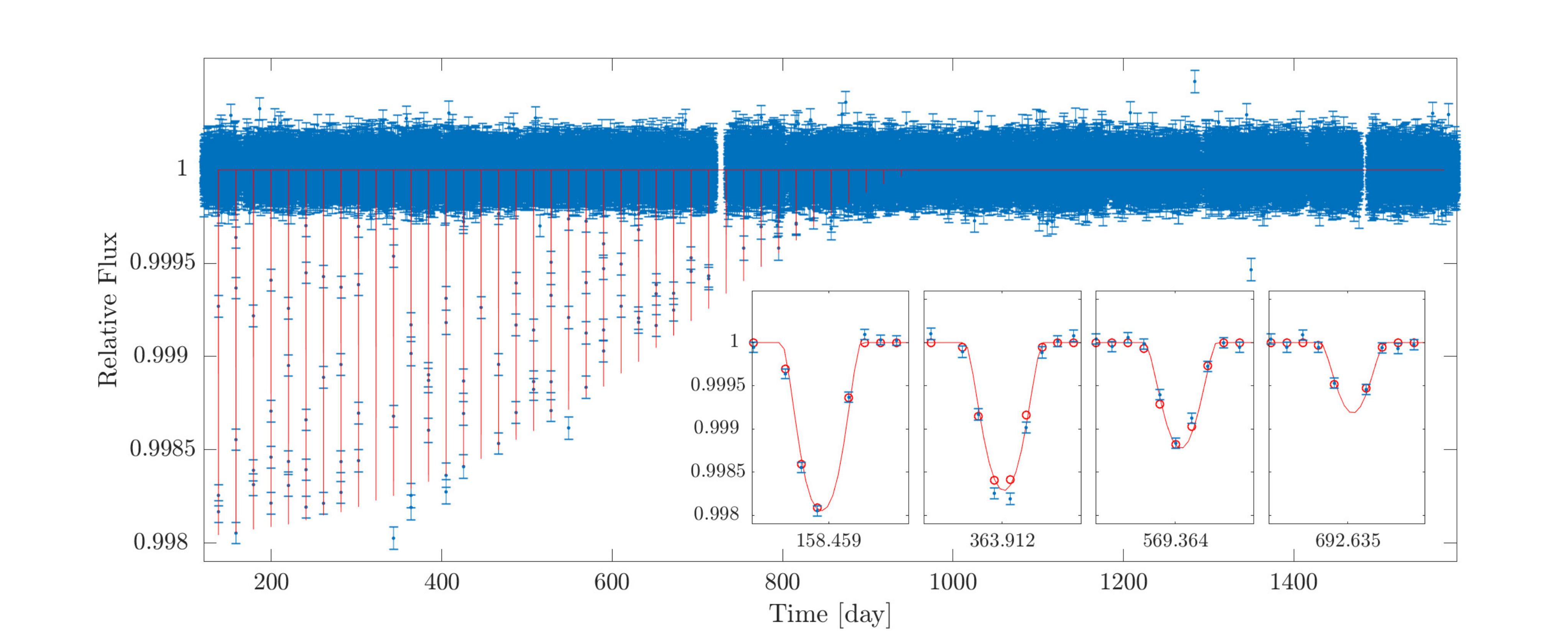}}
    {\caption{The detrended light-curve of KOI 120.01 (in blue) with one of the four scenarios (detailed in \S\ref{LC_Fitting}) model light-curve (in red). Only one of the four models is shown, as they are visually indistinguishable. The transits exhibit an unusual vanishing pattern. On the bottom, several inserts show individual transit events zoomed-in for illustration, each with a single tick at their horizontal axis showing the rough mid-transit times in days of each event. The circles in these panels are the model binned to \textit{Kepler}'s 30min cadence. The vertical span is the same for the large plot and the inserts.}
    \label{fig:KOI120RawLC}}
\end{figure*}

\textbf{Photometry:} \citet{SlawsonEtAl2011} classified it as a detached binary. The light-curve on which this classification was based does not exhibit a secondary eclipse (or primary, if the secondary was actually the one observed) \footnote{http://keplerebs.villanova.edu/overview/?k=11869052}. In principle, an eclipsing binary without a secondary transit is possible in the case of an eccentric, inclined orbit, but this would occur only for a narrow range of geometries. For example, for two Sun-like stars orbiting with an eccentricity of 0.1 and period 20.5 days, the inclination must be in the narrow range of 86.8-87.4$^\circ$ in order for one of the transits to be observed and the other not.
Furthermore, it would also require another massive perturber to torque a binary pair at such a rate that would match the rapidly vanishing observed transits. To conclude, while the possibility of a binary system cannot be ruled out, such a scenario would require some fine tuning of the parameters (literally requiring "the stars to align", so to speak), and should be easily discernible by future observations. Here only the possibility of a planet perturbed by a massive third body is examined below.
 
 \citet{BurkeEtAl2014} and \citet{RoweEtAl2015} defined the status of this planetary candidate as a false positive.
\citet{CoughlinEtAl2014} analyzed \textit{Kepler}'s Q1-Q12 data including KOI 120.01, and defined 685 KOIs as false positives due to contamination of other objects on the CCD. KOI 120.01 was not included in this list of contaminated false positives.
\citet{CampanteEtAl2014} included this object in their catalogue for stellar surface gravity of Sun-like KOI's, estimating astro-seismically its surface gravity to be $\log(g)=4.05$.
\citet{KirkEtAl2016} considered transit depth variations in \textit{Kepler} binary stars. Specifically, they treated KIC 10319590 which exhibits similar vanishing transits, but with a clear secondary eclipse. KOI 120.01 is not included in their list of single-eclipse events.
Other examples of depth changers, which are attributes to eclipsing binaries in eccentric orbits, were treated by \citet{Borkovits2015}. KOI 120.01 is not included in that analysis.
\citet{ArmstrongEtAl2014} supplemented \textit{Kepler}'s photometry with other photometric datasets and fitted temperatures and radii-to-distance ratio to this (assumed) binary among others in their catalog, and estimated the stellar temperatures of the binary members to be $(6233\pm{}350)$K and $(6023\pm{} 540) $K, compatible with Sun-like stars.

As of February 2020, NASA's Exoplanets archive classifies this KOI as a false positive, due to a stellar eclipse flag.

\textbf{Imaging:} Robo-AO images \citep{ZieglerEtAl2017} indicated that within the 4" footprint of a single \textit{Kepler}  pixel there exists another star with a separation of $(1.62 \pm{} 0.06)$~as. The angular separation measurement was later improved by \textit{Gaia} \citep{GAIACollaborationEtAl2018}. The source ID of \textit{Gaia} for these two objects are  2135199457417898240 (brighter object) and  2135199461713461120 (dimmer object) \citep{ZieglerEtAl2018}. The received angular separation from \textit{Gaia} coordinates is $1.5659$", within the error range estimated by ROBO-AO.

\textbf{Spectra:} Spectral measurements by Marcy and Isaacson published on ExoFOP\footnote{https://exofop.ipac.caltech.edu/kepler/} suggest that the source is a  binary. Spectral data of this object was taken by LAMOST survey \citep{ZhaoEtAl2012}, in which the focal plane diameter of each fiber is equivalent to 3.3~as, smaller than \textit{Kepler} pixel field (4~as) but including both Robo-AO/\textit{Gaia} objects. The separation obtained from \textit{Gaia} and ROBO-AO indicates that LAMOST data might include light from both objects, and hence would not resolve them.  

\textbf{Astrometry:} The parallax values taken from \textit{Gaia} data are translated to distances from the solar system of $807 \pm{} 29$~psc (brighter object) and $861 \pm{} 51$~psc (dimmer object). The translation of the angular separation to sky-projected distance between the objects is $\sim$1300~AU. Such a large separation cannot explain a dynamical interaction which would make the 20.5-days-period transit vanish in 800 days. In addition, such a separation cannot account for the RV estimated from the spectral signatures by Marcy and Isaacson mentioned above. Hence, one of the \textit{Gaia} objects 2135199457417898240 or 2135199461713461120 could be an unresolved spectral binary itself.

\subsection{Proposed Scenarios}

The full nature of the transit signals of KOI 120.01 is unclear. There is spectroscopic evidence that the \textit{Kepler} pixel containing KOI 120.01 includes a binary \citep{IsaacsonMarcy2009}, but the radial velocities deduced are at least order of magnitude larger than the $\sim$1~km/s expected for a pair of Sun-like objects orbiting at a minimum separation of $\sim$1300 AU, as indicated by \textit{Gaia}. 
Hence the assumption made here is that one of these two objects is the binary, and the other is a single star. This minimal configuration is required in order to explain all the data with the smallest number of stars.
It is not clear if the two objects resolved by Robo-AO/\textit{Gaia} are gravitationally bound, although this is entirely possible, as according to \textit{Gaia} the two objects have similar proper motions and similar parallaxes. 
A possible explanation for the transit signals is a planet which orbits one of the binary companions, or the single star within the \textit{Kepler} pixel. In all these scenarios, the planet is assumed to gradually move out of transit due to secular precession induced by a massive perturber. As the data displays not only transit timing variations, but also depth and duration variations, 
the proposed dynamical scenario is that the vanishing transit signal is a result of a combined effect of nodal regression and apsidal precession of the planet. We note that already in this minimal scenario the light curve includes the combined effects of four bodies (the occulted star, the transiting object, the torquing object and a distant light-contaminating-only star). Given the current data, models invoking additional planets and degrees of freedom are not justified. Hence, other dynamical scenarios (such as resonant interactions) are not covered in this work.

Four scenarios are examined. Each of the two objects within the Kepler pixel could be the binary, while the other is a single star. The planet could be orbiting the single star, or one of the binary companions. In any case, a simplifying assumption was made - that the brightness of the binary is divided equally between the members, as they are expected to be of similar stellar type due to their matching spectra, similar distance from the Solar System and similar temperatures (\S \ref{subsec:GeneralDescription}). The brightness of the two objects was received from the magnitudes supplied from \textit{Gaia} archive, and the brightness of one of them (assumed to be the binary) was divided equally to two. This approximation limits the analysis to four effective $L_3$ (the total brightness of the background stars relative to the brightness of the transits' host star) values, taken into account for the model light-curve fitting. The four scenarios are summarized in Table \ref{ScenariosTable}.

\begin{table*}[h]
\begin{center}
 \begin{tabular}{||p{1.5cm} |p{2.5cm} |p{2.5cm} |p{2.8cm} |p{2cm}  ||}
  \hline
 Scenario & Brighter Object Type & Dimmer Object Type & Transiting planet host & Effective $L_3$  \\ [0.5ex] 
 \hline\hline
 1 & single & binary & single star & 0.6486  \\
   \hline
   2 & binary & single & single star & 1.5418  \\
   \hline
   3 & binary & single & binary member & 2.2972  \\
   \hline
   4 & single & binary & binary member & 4.0836  \\
   \hline
\end{tabular}
\end{center}
\caption{The four scenarios considered in this work. The object referred to as brighter is \textit{Gaia} 2135199457417898240, and the object referred to as dimmer is \textit{Gaia} 2135199461713461120. It is assumed that one of the objects is actually itself a binary, unresolved in \textit{Gaia} data, but appearing in the spectra. The assumed transiting planet can orbit one member of the binary, or the other (single) star. This results in four scenarios, each with a different effective third light contaminating the transit signals.}
\label{ScenariosTable}
\end{table*}

\subsection{Light-Curve Fitting}
\label{LC_Fitting}
Light-curve is obtained by detrending the SAP flux of the {\it Kepler} data archive using a cosine filter with an iterative outlier removal of data points beyond $5\sigma$. The maximal frequency for the filter corresponded to four times the duration of the first transit, to avoid the transits from affecting the filter.
The fitting of the model light-curve to the detrended data is performed using the parametrization above. For each scenario the third light value is fixed according to Table \ref{ScenariosTable} and used for linearly diluting the transit signal of that model. The non-linear search in the parameters space was performed using DE-MCMC with Snooker update \citep{BraakVrugt2008} with 32 walkers exploring the parameters space. Convergence was reached upon receiving a Gelman-Rubin threshold lower than 1.2 for all model parameters \citep{Gelman1992, Brooks1998} for a sample including 200,000 generations. In addition, it was made sure that the $\chi^2$ improvement drops below the 1$\sigma$ equivalent). Statistical inference of the parameters median and error values was done after removal of the burn-in phase.
Having found a posterior distribution and a best-fitting model, the SAP flux was divided by the best model values in order to remove the transit signals from the raw data, and the light-curve was detrended again. The DE-MCMC process was then repeated for another iteration, beginning at the median values of the former iteration. Each iteration improved the detrending quality.
For the light-curve model calculation, limb-darkening coefficients were taken from NASA exoplanets archive: $u_1=0.3958, u_2=0.2678$. These are fixed and are not fitted for. This approach assumes the stars are of similar type based on the spectra acquired for the binary objects, and thus share the same limb darkening parameters.

\subsection{Photodynamical Fit Results}

The model described in \S\ref{Transit Parametrization} yields precession parameters which describe the three-dimensional orbital motion. For each one of the scenarios in Table \ref{ScenariosTable}, a solution was sought in the full parameter space.
Different solutions in the parameters space were found, having comparable $\chi^2$ values. Therefore, we cannot constrain the model parameters to a single domain. For instance, eccentricity values that range from 0.1 to 0.7 combined with different values of the other parameters could match the data in a comparable quality. 
Hence, we cannot statistically constrain the orbital elements and instead we present below one possible solution for each scenario, to illustrate that the model can explain the data.
The only parameter for which we can learn something from this fitting process is the precession rate. After running multiple DE-MCMC runs, we did get different orbital elements and planet radii that can match the data. However, the precession rate magnitude was always no less than $\sim 1^{\circ} {\rm yr}^{-1}$. We refer to inferences of such precession rates at \S \ref{Discussion}.
The parameters of the selected sample solutions are given in Table \ref{SolutionsTableBestParams}.

Given the number of degrees of freedom, the various models are within $\sim 1 \sigma$ from one another, and therefore are not statistically distinguishable. It is notable that the residuals of the models have an RMS only 15\% higher for in-transit points than out of transit.

\begin{table*}[h]
\begin{center}
 \begin{tabular}{||l|l|l|l|l||}
  \hline
 Scenario & 1 & 2 & 3 & 4  \\ [0.5ex] 
 \hline\hline

$P$ [Day]&$20.5463$&$20.5464$&$20.5506$&$20.5468$\\
     \hline
     $T_{\rm mid}$ [Day]&$137.9002$&$137.9002$&$137.9006$&$137.9012$\\
     \hline
     $R_p/R_*$&$0.115$&$0.0916$&$0.0905$&$0.10516$\\
     \hline
     $a/R_*$&$23.18$&$27.83$&$26.38$&$18.53$\\
     \hline
     $e\sin{\varpi_0}$&$-0.1679$&$-0.1771$&$-0.1843$&$-0.671$\\
     \hline
     $e\cos{\varpi_0}$&$0.1773$&$0.176$&$0.5463$&$0.237$\\
     \hline
     $I\sin{\Omega_0} [^\circ]$&$0.482$&$0.311$&$1.04$&$-0.269$\\
     \hline
     $I\cos{\Omega_0} [^\circ]$&$3.061$&$3.68$&$9.54$&$0.16$\\
     \hline
     $\dot{\Omega} [^\circ {\rm yr}^{-1}]$&$-9.86$&$-9.98$&$-7.18$&$-18.11$\\
     \hline
     $\dot{\omega}/\dot{\Omega}$&$-2$&$-2.0003$&$-2.0062$&$-2$\\
     \hline
     $I_{\rm sky} [^\circ]$&$86.86$&$87.54$&$86.47$&$80.84$\\
     \hline
     $L_3$ (fixed)&$0.6486$&$1.5418$&$2.2972$&$4.0836$\\
     \hline
     $\chi^2$&3139.4506&3144.0894&3149.7334&3157.7414\\
     \hline
     \hline
     $b_0$&1.0232&0.95313&0.91373&0.86982\\
     \hline
     $D_0$ [Hr]&2.4999&2.471&2.4413&2.4262\\
     \hline
     ${\langle \dot{b} \rangle}  [{\rm yr}^{-1}]$&0.040224&0.060789&0.076568&0.10068\\
     \hline

 \end{tabular}
\end{center}
\caption{Parameters values for four solutions fitting the light-curve, each corresponding to one of the scenarios in Table~\ref{ScenariosTable}. The subscript ``0" refers to the value of the parameter at the first mid-transit time (at $\sim$137 days BKJD). The last three rows are quantities derived from the model and not directly fitted - $b_0$ and $D_0$ are the impact parameter and duration of the first transit respectively, and $\langle \dot{b} \rangle$ is the mean change rate of the impact parameter. We note that we intentionally refrain from providing formal errors for the individual parameters since the space of possible solutions for this under-constrained problem, examplified by the four $L_3$ values here, far exceed the formal error ranges.}
\label{SolutionsTableBestParams}
\end{table*}

\begin{figure*}
   {\includegraphics[width=1\linewidth]{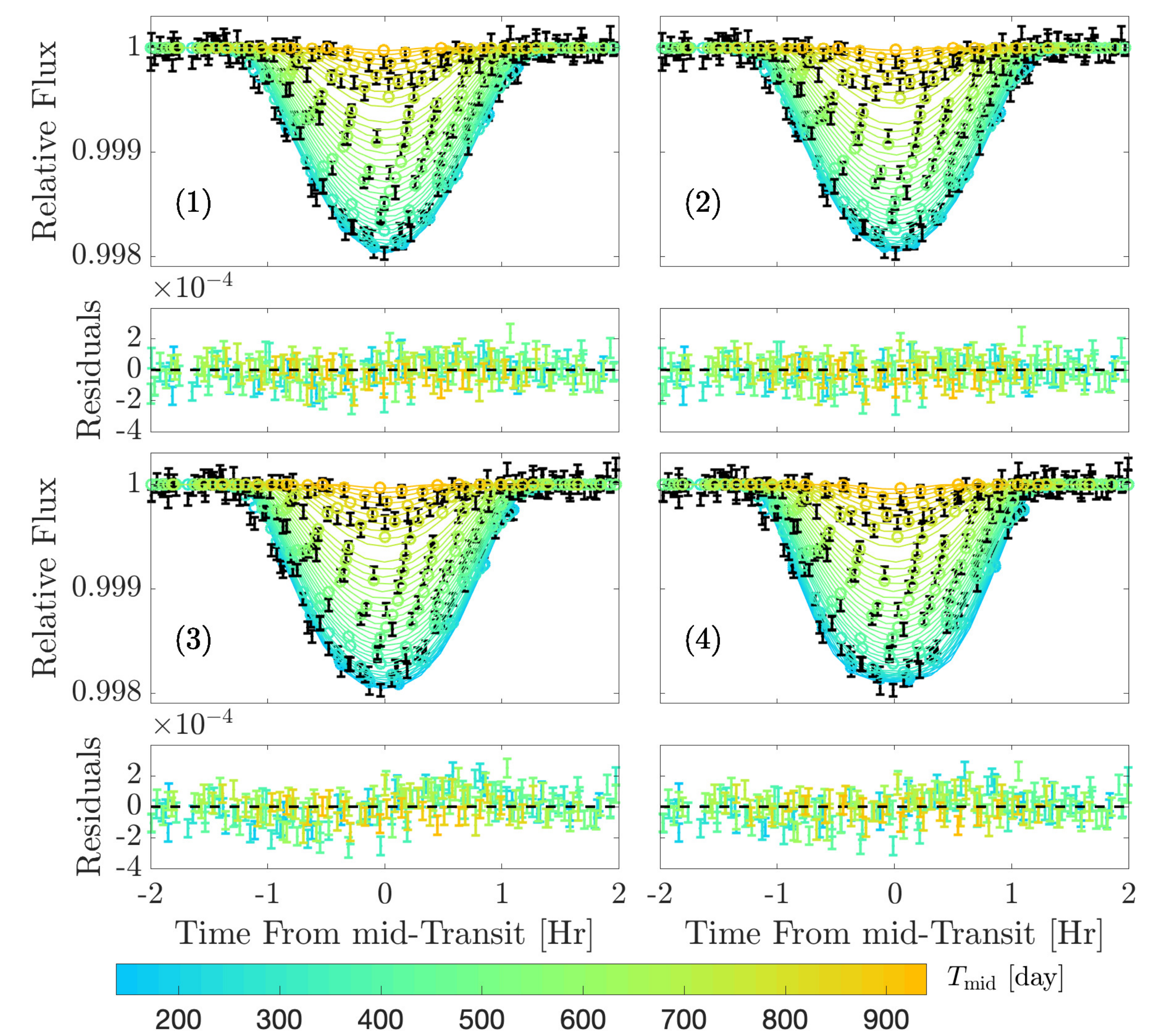}}
   {\caption{Light-curves folded around the different times of mid-transit for each one of the best fit models (the folding is not periodic, it includes the TTV's). Each pair of flux and residuals panels shows the data (black error bars) along with the best fit flux model (solid colored lines) and its binned equivalent (circles). For each plot, the corresponding scenario number is indicated, according to Table~\ref{ScenariosTable}. The color coding represents the time of mid-transit, with warmer colors corresponding to later, shallower, transits. 
   }
    \label{fig:KOI120LC}}
\end{figure*}

\begin{figure}[h]
    {\includegraphics[width=1\linewidth]{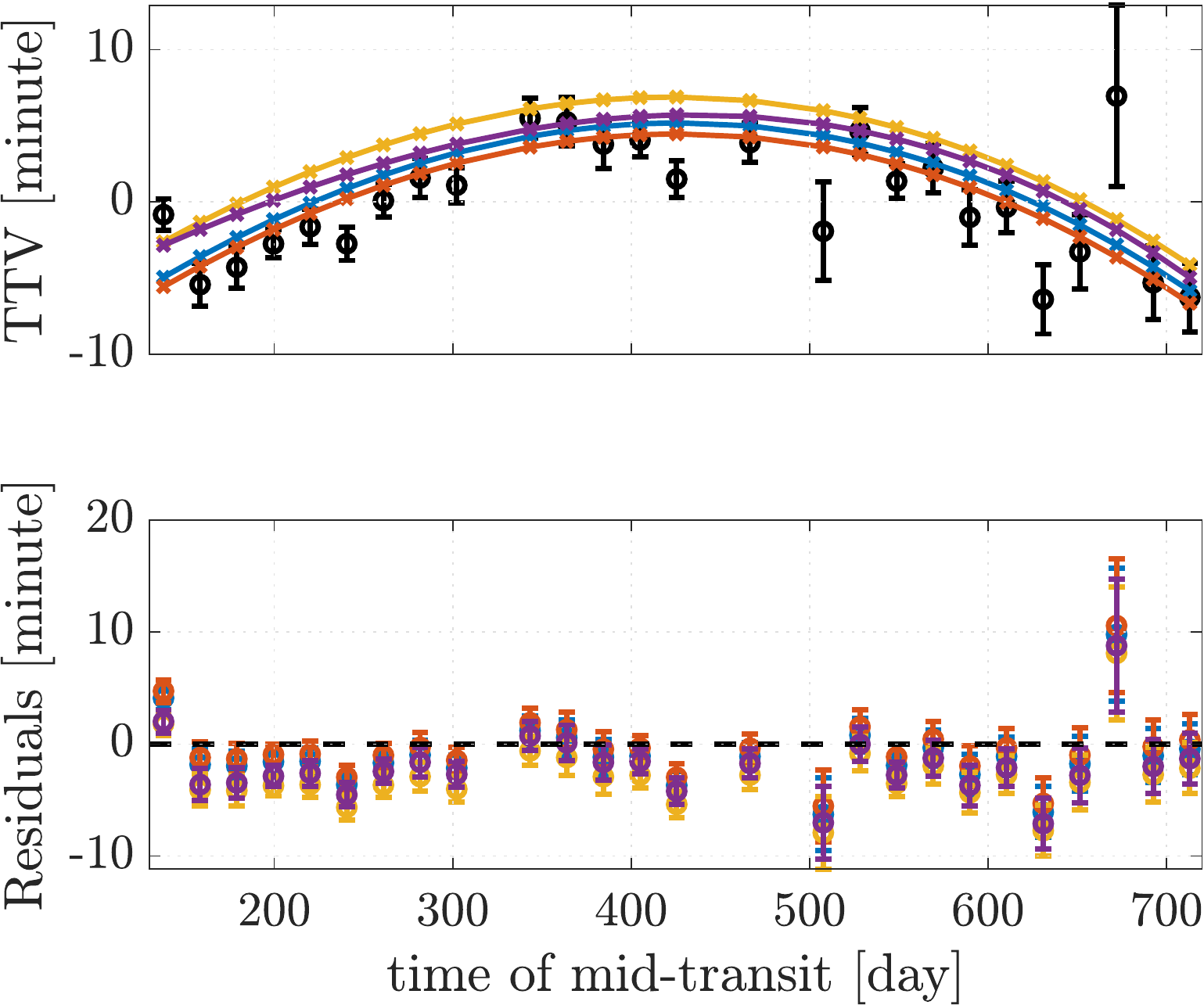}}
    {\caption{\textbf{Top:} TTV's of the four models. The individual times of mid-transit minus their best-fit linear ephemeris are shown above as black errors bars. The shallower the transits become, the larger the errors. The best-fit models of scenarios 1-4 are shown in blue, red, yellow and purple, correspondingly. \textbf{Bottom:} Residuals of the model subtracted from the data.}
    \label{fig:BestFitTTV}}
\end{figure}

\begin{figure*}
    \includegraphics[width=1\linewidth]{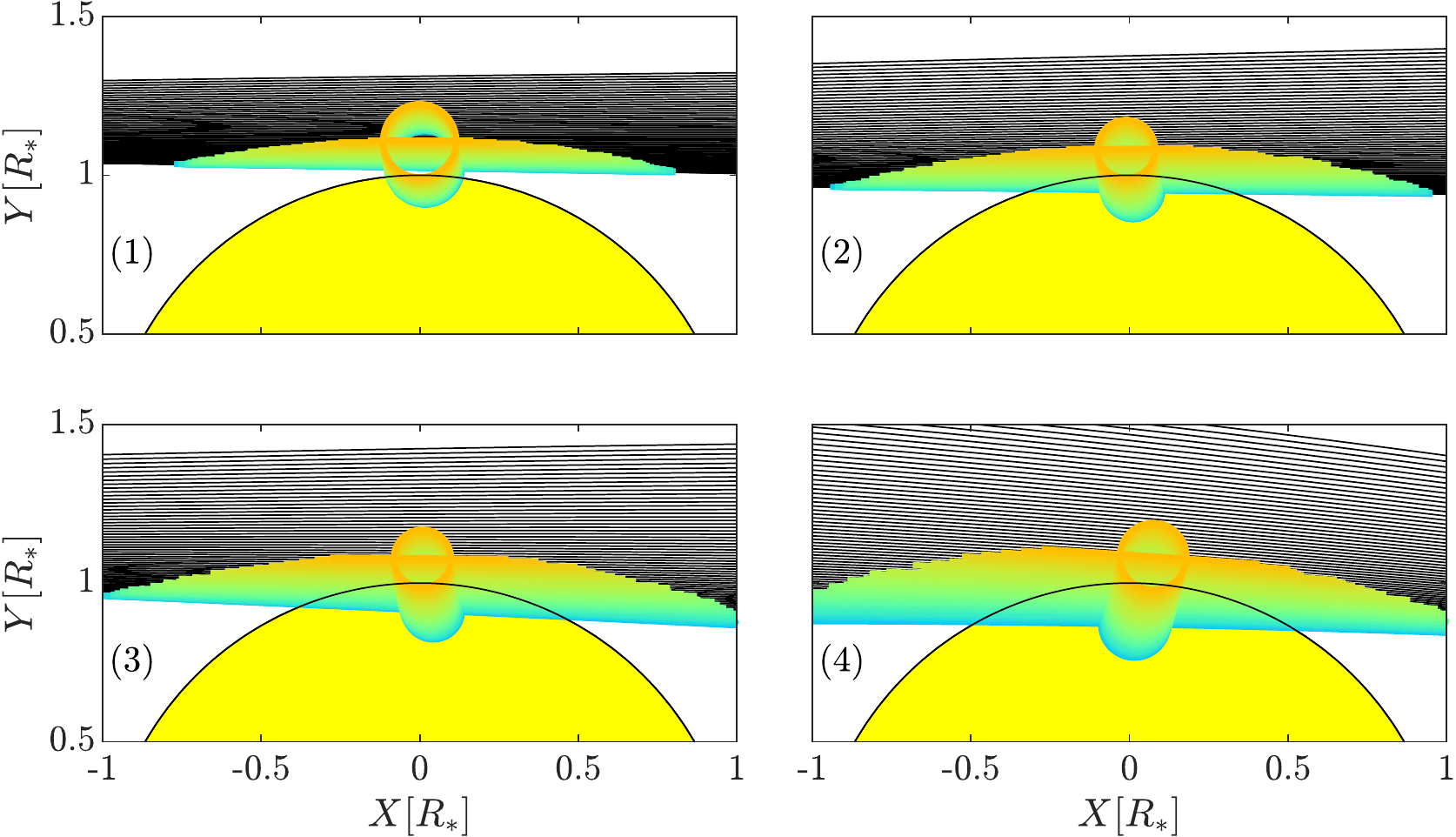}
    \caption{An illustration of the transit chord projected on the sky plane as it moves in time due to the combination of apsidal precession and nodal regression for the four cases discussed in the text. The star disk is represented by the large yellow circle (shown only partially since the image is zoomed in around transit chord). The color coding of the different transit events corresponds to the time of mid-transits, as in the color bar in Figure~\ref{fig:KOI120LC}. Parts of chord that are far away from the transit, or that have no visible transit in that orbit at all, are plotted as black lines. The size of the planet relative to the host star is illustrated by the circle, located at mid-transit point. For each plot, the corresponding scenario number is indicated, according to Table~\ref{ScenariosTable}. In all four cases it is apparent that the mid-transit placement is shifted in time relative to the (fixed) stellar disk - this is due to the three-dimensional effect of the nodal regression.}
    \label{fig:KOI120SkyPlaneZoom}
\end{figure*}

The best fit models, folded around the times of mid-transit, are shown in Figure~\ref{fig:KOI120LC} which emphasized depth variation. Their TTV's are shown in Figure~\ref{fig:BestFitTTV}. The depth variations, resulting from the combination of the apsidal precession and nodal regression, cause the transit chord to move with time. The geometric motion of the chord on the sky plane is illustrated in Figure~\ref{fig:KOI120SkyPlaneZoom}.

Notably, the solutions presented here are only local minima in $\chi^2$ for each of the scenarios; indeed, other local minima do exist that have significantly different physical interpretation but not much worse $\chi^2$ values.


\newpage


\section{Discussion} \label{Discussion}
In the first part of this work a photodynamical analytic approach is presented. Such an approach is useful when the temporal functional dependence of the orbital elements is known and only the magnitude of these effects is sought. Specifically, the model described here is relevant to a scenario in which a distant object gravitationally perturbs an inner transiting planet in a secular manner, and thus the orbital motion is dominated by precession of the periapse and regression of the nodes over the observed epoch. In the second part of the work, this analytic approach is utilized to fit for the vanishing transit signals of KOI 120.01, assumed to arise due to such a configuration. 

It was found that a planet candidate exhibiting both large apsidal precession and large nodal regression rates  can account for the observed transit signal of KOI 120.01, though the phase space is large and multiple solutions are possible given the current data. Four scenarios were treated separately: in each the transit is diluted by a different third light source, corresponding to possible arrangements of the relevant objects within the \textit{Kepler} pixel. Though various sets of parameters were found to fit the data with similar quality, the obtained precession rate magnitude was always in no less than about $1^{\circ} {\rm yr}^{-1}$, and some solutions were found with precession rates as high as $\sim 20^{\circ} {\rm yr}^{-1}$. If indeed the precession scenario describes this transit signal, then the precession rates found here are unusual in their magnitude. For example, the rapidly precessing orbit of KOI 13.01 exhibits a nodal regression rate of about 0.7$^\circ$/year \citep{SzaboEtAl2012}. 

We examine the physical inferences that can be drawn for KOI 120.01 from a scenario of a rapid precession induced by an external unseen companion. Importantly, we note that even when good knowledge of the precession rate is available, the physical properties of the perturber cannot be uniquely constrained, as the observed rate depends on the perturber semi-major axis and mass, and on the mutual inclination between the orbits, in a degenerate manner. The following analysis, therefore, aims to rule out scenarios inconsistent with the data, rather than finding a specific solution.

In order to account for such a precession rate by a low-inclination planetary perturber, a few Jupiter-masses object is required to be located at a semi-major axis such that $\alpha$ is approximately 0.6-0.75, close either to the 2:1 MMR or to the 3:2 MMR. Because the TTV shape is parabolic over the $\sim$780 days during which transits are observed, the super-period of such an interaction must exceed $\sim$ 1500 days. This corresponds to a distance from resonance $\Delta\lesssim0.01$. Such a close proximity to resonance of two giant planets is not common, but possible, for example in the system {\it Kepler}-9 \citep{HolmanEtAl2010, FreudenthalEtAl2018}. However, a close, massive planetary perturber would also induce a chopping signal of order $(\mu'P/2\pi)(P/(P-P'))^2$ \citep{HaddenLithwick2016}, where $\mu'$ and $P'$ are the planet-to-star mass ratio and orbital period of the perturber, respectively. In this case, the chopping signal would be of order of ten minutes, which is not seen in the transiting object's TTVs (Figure~\ref{fig:BestFitTTV}). Therefore, a near-MMR interaction with a giant planet is ruled out by the data. 

Another possibility is that the motion of the planet out of transit does not result from secular interactions, but from other dynamical phenomena which are not covered by our secular model. For example, the planets orbiting the star K2-146 exhibit rapid precession combined with an orbital mean motion resonance, that together yield rapid impact parameter variations \citep{Hamann2019, Lam2019}. Analysis of such scenarios requires adding yet more objects to the model, and is beyond the scope of this study and is left for future work. We focused on scenarios with the minimal number of objects.

In order to explain such a precession rate by a distant perturber located at a few AU (which would yield secular interactions, far from any MMR), its mass would need to be comparable to the host star, {\it i.e.} it would be a binary member. This scenario, if correct, would be notable because of the difficulty in forming giant planets between close binary pair of Sun-like stars. Indeed, planets in S-type orbits were previously found only in binaries separated by more than $\sim5$~AU \citep{GongJi2018}.  
For an S-type planet in a binary system, the question of dynamical stability arises. \citet{HolmanWiegert1999} estimated $a_c$, the critical semi-major axis below which a planet can stably orbit one of the binary members. For a circular orbit of the binary and equal-mass stars, their expression reduces to $a_c=(0.274\pm0.008)a_b$, where $a_b$ is the binary semi-major axis. In our case, if the host star is Sun-like, the orbital period of the presumed planet is translated to a semi-major axis of $\sim0.147$~AU, so for a binary star located at a few AU such a planet is within the stable region. A recent study examined a larger number of orbital configurations for S-type orbits of a planet in a binary system, and reached a similar $a_c=0.26a_b$ for equal-mass binary members at a circular orbit \citep[][fig. 1]{QuarlesEtAl2019}.

\begin{figure*}
    \includegraphics[width=1\linewidth]{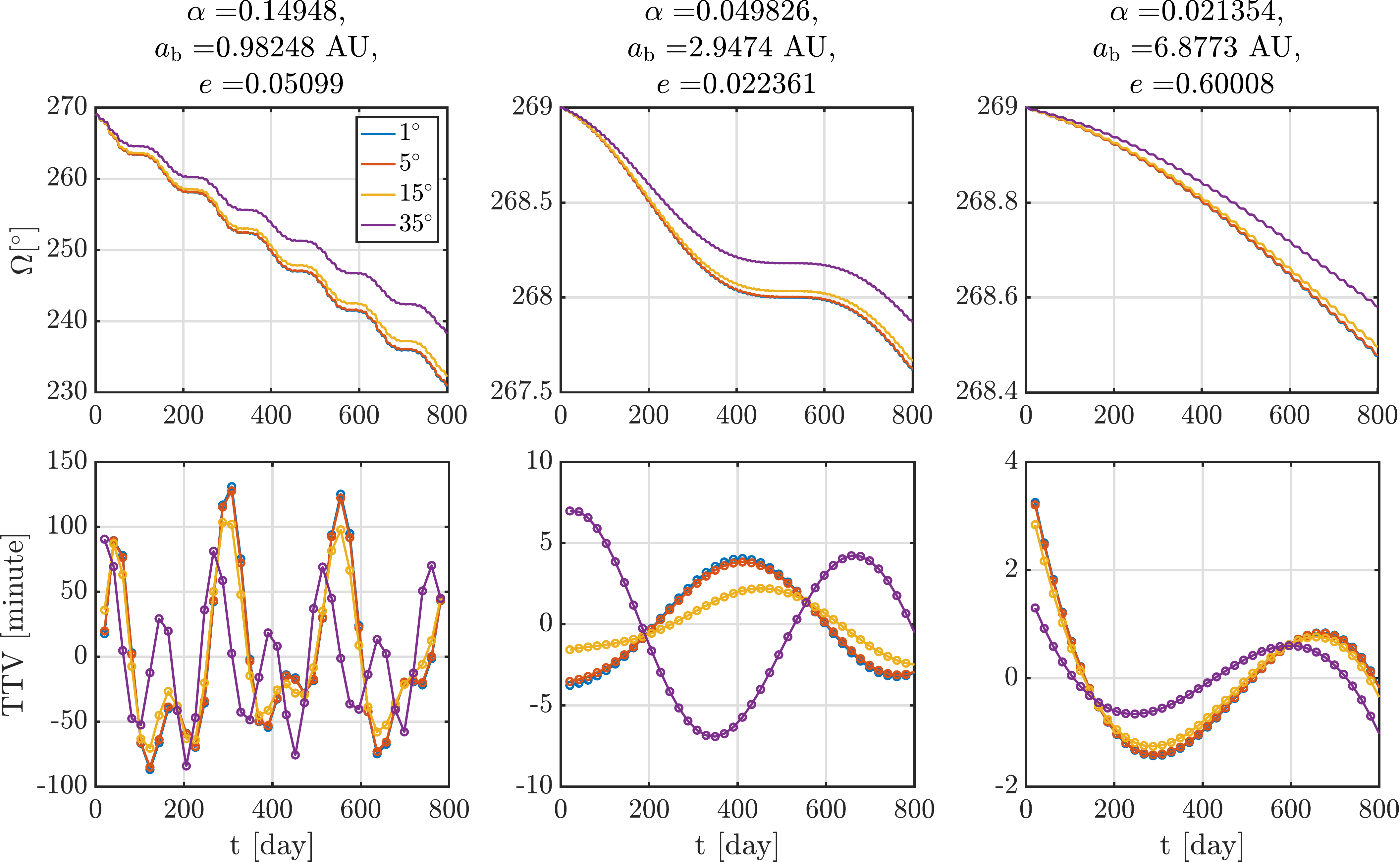}
    \caption{Results of sample N-body integrations, illustrating the effect of a distant stellar perturber on the presumed transiting planet's regression rate and TTV. The analysis is for four possible mutual inclination values between the binary and the planet's orbit: $1^{\circ},5^{\circ},15^{\circ},35^{\circ}$ (see legend). Each configuration is represented by a pair of panels, with a title indicating the values of $\alpha$, $a_{\rm b}$ (semi-major axis of the binary) and $e$ (planet orbital eccentricity). For all cases, the mass of the perturber and the mass of the transit host star are $1M_\odot$.
    Left column: a tight binary at 1~AU generates rapid precession of order 10$^{\circ}$/year (similar to the values obtained by our analytic approach), and a TTV "chopping" pattern (saw-tooth like) at an amplitude of order 100 minutes, not seen in the data. 
    Middle column: a stellar perturber at $\sim3$~AU and a low-eccentricity planet. This configuration yields a precession rate of order $1^{\circ}$/year, within the lower end of the values obtained by our analytic approach, and a TTV pattern of a few minutes, similar in shape and amplitude to that seen in KOI 120.01.
    Right column: a stellar perturber at $\sim 7$~AU and a highly eccentric planet. The stellar binary generates nodal regression rate of order 0.1$^{\circ}$/year (lower than our analytic model values), and a TTV pattern of order a few minutes (similar to the TTVs seen in the data).
    Generally, the TTV amplitude  grows with the eccentricity of the transiting planets, and shrinks with the binary separation, so their combination can match the observed value.}
    \label{fig:SampleTTVAndPrecession}
\end{figure*}

In order to check the feasibility of a configuration of an S-type planet, we ran a several N-body integrations of a scenario in which the perturber is a binary of the same mass as the host star, located at a few AU. We find that the precession rate matches the theoretical secular values well, but the TTV's might be larger than the observed ones. We note that the secular model used above takes into account only the TTV part that arises from the orbital precession, but not other dynamical effects which become more and more significant at higher eccentricity values of the transiting planet. For this reason, the secular model underestimates the TTV, and is prone to overestimate the eccentricity, or overestimate the precession rate required to explain the data. This may explain the high eccentricity and precession rates values of the presumed transiting planet obtained by using the secular precession model.
The explanation above is illustrated in Fig. \ref{fig:SampleTTVAndPrecession}, which shows sample TTV patterns and nodal regression rates for a few specific scenarios. Specifically, it shows that the scenario of secular precession combined with a parabolic TTV patterns can occur for an S-type planet perturbed by the binary member, and hence qualitatively, the configuration of an S-type planet perturbed by a binary member is possible for KOI 120.01, if the binary separation is large enough. 

We conclude that the simple precession model utilized here yields a decent fit to the photometry (\S \ref{LC_Fitting}), but the physical inferences derived from the obtained solutions, along with experimenting with N-body simulations, imply that the dynamical nature of the system is likely more complex. Full N-body models (and probably more observational data) would be required in order to constrain the characteristics of the perturber and the transiting object.

The transit signal may still arise from an eclipsing binary, itself perturbed by a tertiary massive star. If this is the case, the orbit should be inclined and eccentric as no secondary eclipse is visible.


\section{Prospects for Future Observations}

A possible clue that can guide future observations is available from the detection of the motion of the center of light on the {\it Kepler} CCD.
For two objects of a comparable apparent magnitude, and one partially occulted with a relative transit depth of $\sim2\times10^{-3}$, the expected motion of the center of light due to the transit is $\sim1\times10^{-3}$ of their separation. Such a motion is indeed reported in the {\it Kepler} database\footnote{https://exoplanetarchive.ipac.caltech.edu/data/ KeplerData/011/011869/011869052/dv/kplr011869052-20160209194854\_dvr.pdf}. According to the {\it Kepler} centroid analysis, examining the in-transit points, the centroid RA increases by $\sim$2~mas and the DEC decreases by $\sim$1~mas. {\it Gaia} data places object 2135199457417898240 at a separation from 2135199461713461120 of -1.884~as in RA and 0.9978~as in DEC, matching the {\it Kepler} observed centroid shift if the transits are of 2135199457417898240. 
To emphasize, both the magnitude and direction of the centroid shift are in agreement with predictions based on {\it Gaia} astrometry, lending significant support to {\it Gaia} 2135199457417898240 as the transits host, and to estimating that the perturber itself does not contribute significant amount of light.




Followup RV measurements dedicated to finding this deduced companion star and the observed transiting object would need to be able to resolve two objects separated by 1.56~as. As the orbital period of the presumed planet is about 20.5 days, the RV semi-amplitude arising from it should be of order a few to a few tens of m/s, achievable with the observational capabilities existing today. The RV semi-amplitude arising from the binary member would be much larger. Follow-up observations, including RV measurements, are encouraged in order to shed more light on the nature of this unique \textit{Kepler} object, and determine if indeed it harbours an unusual planetary system.

\acknowledgments
We acknowledge helpful discussions with Daniel Fabrycky, Eric Ford and Re'em Sari and thank an anonymous reviewer for their constructive comments.  This study was supported by the Helen Kimmel Center for Planetary Sciences and the Minerva Center for Life Under Extreme Planetary Conditions \#13599 at the Weizmann Institute of Science.
\newpage

\bibliography{Mybib.bib}{}

\begin{thebibliography}{}
\expandafter\ifx\csname natexlab\endcsname\relax\def\natexlab#1{#1}\fi
\providecommand{\url}[1]{\href{#1}{#1}}
\providecommand{\dodoi}[1]{doi:~\href{http://doi.org/#1}{\nolinkurl{#1}}}
\providecommand{\doeprint}[1]{\href{http://ascl.net/#1}{\nolinkurl{http://ascl.net/#1}}}
\providecommand{\doarXiv}[1]{\href{https://arxiv.org/abs/#1}{\nolinkurl{https://arxiv.org/abs/#1}}}

\bibitem[{{Agol} {et~al.}(2005){Agol}, {Steffen}, {Sari}, \&
  {Clarkson}}]{AgolSteffenSariClarkson2005}
{Agol}, E., {Steffen}, J., {Sari}, R., \& {Clarkson}, W. 2005, \mnras, 359,
  567, \dodoi{10.1111/j.1365-2966.2005.08922.x}

\bibitem[{{Armstrong} {et~al.}(2014){Armstrong}, {G{\'o}mez Maqueo Chew},
  {Faedi}, \& {Pollacco}}]{ArmstrongEtAl2014}
{Armstrong}, D.~J., {G{\'o}mez Maqueo Chew}, Y., {Faedi}, F., \& {Pollacco}, D.
  2014, \mnras, 437, 3473, \dodoi{10.1093/mnras/stt2146}

\bibitem[{{Ballard} \& {Johnson}(2016)}]{BallardJohnson2016}
{Ballard}, S., \& {Johnson}, J.~A. 2016, \apj, 816, 66,
  \dodoi{10.3847/0004-637X/816/2/66}

\bibitem[{{Batalha} {et~al.}(2013){Batalha}, {Rowe}, {Bryson}, {Barclay},
  {Burke}, {Caldwell}, {Christiansen}, {Mullally}, {Thompson}, {Brown},
  {Dupree}, {Fabrycky}, {Ford}, {Fortney}, {Gilliland}, {Isaacson}, {Latham},
  {Marcy}, {Quinn}, {Ragozzine}, {Shporer}, {Borucki}, {Ciardi}, {Gautier},
  {Haas}, {Jenkins}, {Koch}, {Lissauer}, {Rapin}, {Basri}, {Boss}, {Buchhave},
  {Carter}, {Charbonneau}, {Christensen-Dalsgaard}, {Clarke}, {Cochran},
  {Demory}, {Desert}, {Devore}, {Doyle}, {Esquerdo}, {Everett}, {Fressin},
  {Geary}, {Girouard}, {Gould}, {Hall}, {Holman}, {Howard}, {Howell},
  {Ibrahim}, {Kinemuchi}, {Kjeldsen}, {Klaus}, {Li}, {Lucas}, {Meibom},
  {Morris}, {Pr{\v{s}}a}, {Quintana}, {Sanderfer}, {Sasselov}, {Seader},
  {Smith}, {Steffen}, {Still}, {Stumpe}, {Tarter}, {Tenenbaum}, {Torres},
  {Twicken}, {Uddin}, {Van Cleve}, {Walkowicz}, \& {Welsh}}]{BatalhaEtAl2013}
{Batalha}, N.~M., {Rowe}, J.~F., {Bryson}, S.~T., {et~al.} 2013, The
  Astrophysical Journal Supplement Series, 204, 24,
  \dodoi{10.1088/0067-0049/204/2/24}

\bibitem[{{Blanchet} {et~al.}(2019){Blanchet}, {H{\'e}brard}, \&
  {Larrouturou}}]{BlanchetEtAl2019}
{Blanchet}, L., {H{\'e}brard}, G., \& {Larrouturou}, F. 2019, arXiv e-prints,
  arXiv:1905.06630.
\newblock \doarXiv{1905.06630}

\bibitem[{{Borkovits} {et~al.}(2015){Borkovits}, {Rappaport}, {Hajdu}, \&
  {Sztakovics}}]{Borkovits2015}
{Borkovits}, T., {Rappaport}, S., {Hajdu}, T., \& {Sztakovics}, J. 2015,
  \mnras, 448, 946, \dodoi{10.1093/mnras/stv015}

\bibitem[{{Borucki} {et~al.}(2010){Borucki}, {Koch}, {Basri}, {Batalha},
  {Brown}, {Caldwell}, {Caldwell}, {Christensen-Dalsgaard}, {Cochran},
  {DeVore}, {Dunham}, {Dupree}, {Gautier}, {Geary}, {Gilliland}, {Gould},
  {Howell}, {Jenkins}, {Kondo}, {Latham}, {Marcy}, {Meibom}, {Kjeldsen},
  {Lissauer}, {Monet}, {Morrison}, {Sasselov}, {Tarter}, {Boss}, {Brownlee},
  {Owen}, {Buzasi}, {Charbonneau}, {Doyle}, {Fortney}, {Ford}, {Holman},
  {Seager}, {Steffen}, {Welsh}, {Rowe}, {Anderson}, {Buchhave}, {Ciardi},
  {Walkowicz}, {Sherry}, {Horch}, {Isaacson}, {Everett}, {Fischer}, {Torres},
  {Johnson}, {Endl}, {MacQueen}, {Bryson}, {Dotson}, {Haas}, {Kolodziejczak},
  {Van Cleve}, {Chandrasekaran}, {Twicken}, {Quintana}, {Clarke}, {Allen},
  {Li}, {Wu}, {Tenenbaum}, {Verner}, {Bruhweiler}, {Barnes}, \&
  {Prsa}}]{BoruckiEtAl2010}
{Borucki}, W.~J., {Koch}, D., {Basri}, G., {et~al.} 2010, Science, 327, 977,
  \dodoi{10.1126/science.1185402}

\bibitem[{Brooks \& Gelman(1998)}]{Brooks1998}
Brooks, S.~P., \& Gelman, A. 1998, Journal of Computational and Graphical
  Statistics, 7, 434, \dodoi{10.1080/10618600.1998.10474787}

\bibitem[{{Burke} {et~al.}(2014){Burke}, {Bryson}, {Mullally}, {Rowe},
  {Christiansen}, {Thompson}, {Coughlin}, {Haas}, {Batalha}, {Caldwell},
  {Jenkins}, {Still}, {Barclay}, {Borucki}, {Chaplin}, {Ciardi}, {Clarke},
  {Cochran}, {Demory}, {Esquerdo}, {Gautier}, {Gilliland}, {Girouard}, {Havel},
  {Henze}, {Howell}, {Huber}, {Latham}, {Li}, {Morehead}, {Morton}, {Pepper},
  {Quintana}, {Ragozzine}, {Seader}, {Shah}, {Shporer}, {Tenenbaum}, {Twicken},
  \& {Wolfgang}}]{BurkeEtAl2014}
{Burke}, C.~J., {Bryson}, S.~T., {Mullally}, F., {et~al.} 2014, \apjs, 210, 19,
  \dodoi{10.1088/0067-0049/210/2/19}

\bibitem[{{Campante} {et~al.}(2014){Campante}, {Chaplin}, {Lund}, {Huber},
  {Hekker}, {Garc{\'{\i}}a}, {Corsaro}, {Handberg}, {Miglio}, {Arentoft},
  {Basu}, {Bedding}, {Christensen-Dalsgaard}, {Davies}, {Elsworth},
  {Gilliland}, {Karoff}, {Kawaler}, {Kjeldsen}, {Lundkvist}, {Metcalfe}, {Silva
  Aguirre}, \& {Stello}}]{CampanteEtAl2014}
{Campante}, T.~L., {Chaplin}, W.~J., {Lund}, M.~N., {et~al.} 2014, \apj, 783,
  123, \dodoi{10.1088/0004-637X/783/2/123}

\bibitem[{{Coughlin} {et~al.}(2014){Coughlin}, {Thompson}, {Bryson}, {Burke},
  {Caldwell}, {Christiansen}, {Haas}, {Howell}, {Jenkins}, {Kolodziejczak},
  {Mullally}, \& {Rowe}}]{CoughlinEtAl2014}
{Coughlin}, J.~L., {Thompson}, S.~E., {Bryson}, S.~T., {et~al.} 2014, \aj, 147,
  119, \dodoi{10.1088/0004-6256/147/5/119}

\bibitem[{{Einstein}(1916)}]{Einstein1916}
{Einstein}, A. 1916, Annalen der Physik, 354, 769,
  \dodoi{10.1002/andp.19163540702}

\bibitem[{{Fabrycky} {et~al.}(2014){Fabrycky}, {Lissauer}, {Ragozzine}, {Rowe},
  {Steffen}, {Agol}, {Barclay}, {Batalha}, {Borucki}, {Ciardi}, {Ford},
  {Gautier}, {Geary}, {Holman}, {Jenkins}, {Li}, {Morehead}, {Morris},
  {Shporer}, {Smith}, {Still}, \& {Van Cleve}}]{FabryckyEtAl2014}
{Fabrycky}, D.~C., {Lissauer}, J.~J., {Ragozzine}, D., {et~al.} 2014, \apj,
  790, 146, \dodoi{10.1088/0004-637X/790/2/146}

\bibitem[{{Freudenthal} {et~al.}(2018){Freudenthal}, {von Essen}, {Dreizler},
  {Wedemeyer}, {Agol}, {Morris}, {Becker}, {Mallonn}, {Hoyer}, {Ofir}, {Tal-
  Or}, {Deeg}, {Herrero}, {Ribas}, {Khalafinejad}, {Hern{\'a}ndez}, \&
  {Rodr{\'\i}guez S.}}]{FreudenthalEtAl2018}
{Freudenthal}, J., {von Essen}, C., {Dreizler}, S., {et~al.} 2018, \aap, 618,
  A41, \dodoi{10.1051/0004-6361/201833436}

\bibitem[{{Gaia Collaboration} {et~al.}(2018){Gaia Collaboration}, {Brown},
  {Vallenari}, {Prusti}, {de Bruijne}, {Babusiaux}, \&
  {Bailer-Jones}}]{GAIACollaborationEtAl2018}
{Gaia Collaboration}, {Brown}, A.~G.~A., {Vallenari}, A., {et~al.} 2018, ArXiv
  e-prints.
\newblock \doarXiv{1804.09365}

\bibitem[{Gelman \& Rubin(1992)}]{Gelman1992}
Gelman, A., \& Rubin, D.~B. 1992, Statist. Sci., 7, 457,
  \dodoi{10.1214/ss/1177011136}

\bibitem[{{Gong} \& {Ji}(2018)}]{GongJi2018}
{Gong}, Y.-X., \& {Ji}, J. 2018, \mnras, 478, 4565,
  \dodoi{10.1093/mnras/sty1300}

\bibitem[{{Hadden} \& {Lithwick}(2016)}]{HaddenLithwick2016}
{Hadden}, S., \& {Lithwick}, Y. 2016, \apj, 828, 44,
  \dodoi{10.3847/0004-637X/828/1/44}

\bibitem[{{Hamann} {et~al.}(2019){Hamann}, {Montet}, {Fabrycky}, {Agol}, \&
  {Kruse}}]{Hamann2019}
{Hamann}, A., {Montet}, B.~T., {Fabrycky}, D.~C., {Agol}, E., \& {Kruse}, E.
  2019, arXiv e-prints, arXiv:1907.10620.
\newblock \doarXiv{1907.10620}

\bibitem[{{He} {et~al.}(2019){He}, {Ford}, \& {Ragozzine}}]{Matthias2019}
{He}, M.~Y., {Ford}, E.~B., \& {Ragozzine}, D. 2019, \mnras, 490, 4575,
  \dodoi{10.1093/mnras/stz2869}

\bibitem[{{Holman} \& {Murray}(2005)}]{HolmanMurray2005}
{Holman}, M.~J., \& {Murray}, N.~W. 2005, Science, 307, 1288,
  \dodoi{10.1126/science.1107822}

\bibitem[{{Holman} \& {Wiegert}(1999)}]{HolmanWiegert1999}
{Holman}, M.~J., \& {Wiegert}, P.~A. 1999, \aj, 117, 621,
  \dodoi{10.1086/300695}

\bibitem[{{Holman} {et~al.}(2010){Holman}, {Fabrycky}, {Ragozzine}, {Ford},
  {Steffen}, {Welsh}, {Lissauer}, {Latham}, {Marcy}, {Walkowicz}, {Batalha},
  {Jenkins}, {Rowe}, {Cochran}, {Fressin}, {Torres}, {Buchhave}, {Sasselov},
  {Borucki}, {Koch}, {Basri}, {Brown}, {Caldwell}, {Charbonneau}, {Dunham},
  {Gautier}, {Geary}, {Gilliland}, {Haas}, {Howell}, {Ciardi}, {Endl},
  {Fischer}, {F{\"u}r{\'e}sz}, {Hartman}, {Isaacson}, {Johnson}, {MacQueen},
  {Moorhead}, {Morehead}, \& {Orosz}}]{HolmanEtAl2010}
{Holman}, M.~J., {Fabrycky}, D.~C., {Ragozzine}, D., {et~al.} 2010, Science,
  330, 51, \dodoi{10.1126/science.1195778}

\bibitem[{Isaacson \& Marcy(2009)}]{IsaacsonMarcy2009}
Isaacson, H., \& Marcy, G. 2009

\bibitem[{{Kirk} {et~al.}(2016){Kirk}, {Conroy}, {Pr{\v s}a}, {Abdul-Masih},
  {Kochoska}, {Matijevi{\v c}}, {Hambleton}, {Barclay}, {Bloemen}, {Boyajian},
  {Doyle}, {Fulton}, {Hoekstra}, {Jek}, {Kane}, {Kostov}, {Latham}, {Mazeh},
  {Orosz}, {Pepper}, {Quarles}, {Ragozzine}, {Shporer}, {Southworth},
  {Stassun}, {Thompson}, {Welsh}, {Agol}, {Derekas}, {Devor}, {Fischer},
  {Green}, {Gropp}, {Jacobs}, {Johnston}, {LaCourse}, {Saetre}, {Schwengeler},
  {Toczyski}, {Werner}, {Garrett}, {Gore}, {Martinez}, {Spitzer}, {Stevick},
  {Thomadis}, {Vrijmoet}, {Yenawine}, {Batalha}, \& {Borucki}}]{KirkEtAl2016}
{Kirk}, B., {Conroy}, K., {Pr{\v s}a}, A., {et~al.} 2016, \aj, 151, 68,
  \dodoi{10.3847/0004-6256/151/3/68}

\bibitem[{{Lai} \& {Pu}(2017)}]{LaiPu2017}
{Lai}, D., \& {Pu}, B. 2017, \aj, 153, 42, \dodoi{10.3847/1538-3881/153/1/42}

\bibitem[{Lam {et~al.}(2019)Lam, Korth, Masuda, Csizmadia, Eigm\"{u}ller,
  Stef\'{a}nsson, Endl, Albrecht, Luque, Livingston, Hirano, Sobrino,
  Barrag\'{a}n, Cabrera, Carleo, Chaushev, Cochran, Dai, Leon, Deeg, Erikson,
  Esposito, Fridlund, Fukui, Gandolfi, Georgieva, Cuesta, Grziwa, Guenther,
  Hatzes, Hidalgo, Hjorth, Kabath, Knudstrup, Lund, Mahadevan, Mathur,
  Rodr\'{i}guez, Murgas, Narita, Nespral, Niraula, Palle, P\"{a}tzold, Persson,
  Prieto-Arranz, Rauer, Redfield, Ribas, Robertson, Skarka, Smith, Subjak, \&
  Eylen}]{Lam2019}
Lam, K. W.~F., Korth, J., Masuda, K., {et~al.} 2019, arXiv:1907.11141

\bibitem[{{Lee} \& {Peale}(2003)}]{LeePeale2003}
{Lee}, M.~H., \& {Peale}, S.~J. 2003, \apj, 592, 1201, \dodoi{10.1086/375857}

\bibitem[{{Lissauer} {et~al.}(2011){Lissauer}, {Ragozzine}, {Fabrycky},
  {Steffen}, {Ford}, {Jenkins}, {Shporer}, {Holman}, {Rowe}, {Quintana},
  {Batalha}, {Borucki}, {Bryson}, {Caldwell}, {Carter}, {Ciardi}, {Dunham},
  {Fortney}, {Gautier}, {Howell}, {Koch}, {Latham}, {Marcy}, {Morehead}, \&
  {Sasselov}}]{Lissauer2011}
{Lissauer}, J.~J., {Ragozzine}, D., {Fabrycky}, D.~C., {et~al.} 2011, The
  Astrophysical Journal Supplement Series, 197, 8,
  \dodoi{10.1088/0067-0049/197/1/8}

\bibitem[{{Mandel} \& {Agol}(2002)}]{MandelAgol2002}
{Mandel}, K., \& {Agol}, E. 2002, \apj, 580, L171, \dodoi{10.1086/345520}

\bibitem[{{Mills} {et~al.}(2019){Mills}, {Howard}, {Weiss}, {Steffen},
  {Isaacson}, {Fulton}, {Petigura}, {Kosiarek}, {Hirsch}, \&
  {Boisvert}}]{MillsEtAl2019}
{Mills}, S.~M., {Howard}, A.~W., {Weiss}, L.~M., {et~al.} 2019, \aj, 157, 145,
  \dodoi{10.3847/1538-3881/ab0899}

\bibitem[{{Miralda-Escud{\'e}}(2002)}]{MiraldaEscude2002}
{Miralda-Escud{\'e}}, J. 2002, \apj, 564, 1019

\bibitem[{{Moro-Mart{\'\i}n} {et~al.}(2007){Moro-Mart{\'\i}n}, {Malhotra},
  {Carpenter}, {Hillenbrand}, {Wolf}, {Meyer}, {Hollenbach}, {Najita}, \&
  {Henning}}]{MoroMartinEtAl2007}
{Moro-Mart{\'\i}n}, A., {Malhotra}, R., {Carpenter}, J.~M., {et~al.} 2007,
  \apj, 668, 1165, \dodoi{10.1086/521093}

\bibitem[{{Murray} \& {Dermott}(1999)}]{SSD1999}
{Murray}, C.~D., \& {Dermott}, S.~F. 1999, {Solar system dynamics}

\bibitem[{{Quarles} {et~al.}(2019){Quarles}, {Li}, {Kostov}, \&
  {Haghighipour}}]{QuarlesEtAl2019}
{Quarles}, B., {Li}, G., {Kostov}, V., \& {Haghighipour}, N. 2019, arXiv
  e-prints, arXiv:1912.11019.
\newblock \doarXiv{1912.11019}

\bibitem[{{Rowe} {et~al.}(2015){Rowe}, {Coughlin}, {Antoci}, {Barclay},
  {Batalha}, {Borucki}, {Burke}, {Bryson}, {Caldwell}, {Campbell},
  {Catanzarite}, {Christiansen}, {Cochran}, {Gilliland}, {Girouard}, {Haas},
  {He{\l}miniak}, {Henze}, {Hoffman}, {Howell}, {Huber}, {Hunter},
  {Jang-Condell}, {Jenkins}, {Klaus}, {Latham}, {Li}, {Lissauer}, {McCauliff},
  {Morris}, {Mullally}, {Ofir}, {Quarles}, {Quintana}, {Sabale}, {Seader},
  {Shporer}, {Smith}, {Steffen}, {Still}, {Tenenbaum}, {Thompson}, {Twicken},
  {Van Laerhoven}, {Wolfgang}, \& {Zamudio}}]{RoweEtAl2015}
{Rowe}, J.~F., {Coughlin}, J.~L., {Antoci}, V., {et~al.} 2015, \apjs, 217, 16,
  \dodoi{10.1088/0067-0049/217/1/16}

\bibitem[{{Slawson} {et~al.}(2011){Slawson}, {Pr{\v s}a}, {Welsh}, {Orosz},
  {Rucker}, {Batalha}, {Doyle}, {Engle}, {Conroy}, {Coughlin}, {Gregg},
  {Fetherolf}, {Short}, {Windmiller}, {Fabrycky}, {Howell}, {Jenkins}, {Uddin},
  {Mullally}, {Seader}, {Thompson}, {Sanderfer}, {Borucki}, \&
  {Koch}}]{SlawsonEtAl2011}
{Slawson}, R.~W., {Pr{\v s}a}, A., {Welsh}, W.~F., {et~al.} 2011, \aj, 142,
  160, \dodoi{10.1088/0004-6256/142/5/160}

\bibitem[{{Szab{\'o}} {et~al.}(2012){Szab{\'o}}, {P{\'a}l}, {Derekas}, {Simon},
  {Szalai}, \& {Kiss}}]{SzaboEtAl2012}
{Szab{\'o}}, G.~M., {P{\'a}l}, A., {Derekas}, A., {et~al.} 2012, \mnras, 421,
  L122, \dodoi{10.1111/j.1745-3933.2012.01219.x}

\bibitem[{ter Braak \& Vrugt(2008)}]{BraakVrugt2008}
ter Braak, C. J.~F., \& Vrugt, J.~A. 2008, Statistics and Computing, 18, 435

\bibitem[{{Wu} \& {Goldreich}(2002)}]{WuGoldreich2002}
{Wu}, Y., \& {Goldreich}, P. 2002, \apj, 564, 1024, \dodoi{10.1086/324193}

\bibitem[{{Xie} {et~al.}(2016){Xie}, {Dong}, {Zhu}, {Huber}, {Zheng}, {De Cat},
  {Fu}, {Liu}, {Luo}, \& {Wu}}]{XieEtAl2016}
{Xie}, J.-W., {Dong}, S., {Zhu}, Z., {et~al.} 2016, Proceedings of the National
  Academy of Science, 113, 11431, \dodoi{10.1073/pnas.1604692113}

\bibitem[{{Xu} \& {Fabrycky}(2019)}]{XuFabrycky2019}
{Xu}, W., \& {Fabrycky}, D. 2019, arXiv e-prints, arXiv:1904.02290.
\newblock \doarXiv{1904.02290}

\bibitem[{{Zhao} {et~al.}(2012){Zhao}, {Zhao}, {Chu}, {Jing}, \&
  {Deng}}]{ZhaoEtAl2012}
{Zhao}, G., {Zhao}, Y.-H., {Chu}, Y.-Q., {Jing}, Y.-P., \& {Deng}, L.-C. 2012,
  Research in Astronomy and Astrophysics, 12, 723,
  \dodoi{10.1088/1674-4527/12/7/002}

\bibitem[{{Ziegler} {et~al.}(2018){Ziegler}, {Law}, {Baranec}, {Morton},
  {Riddle}, {Huber}, {Mahadevan}, \& {Pepper}}]{ZieglerEtAl2018}
{Ziegler}, C., {Law}, N.~M., {Baranec}, C., {et~al.} 2018, ArXiv e-prints.
\newblock \doarXiv{1806.10142}

\bibitem[{{Ziegler} {et~al.}(2017){Ziegler}, {Law}, {Baranec}, {Riddle},
  {Duev}, {Howard}, {Jensen-Clem}, {Kulkarni}, {Morton}, \&
  {Salama}}]{ZieglerEtAl2017}
---. 2017, ArXiv e-prints.
\newblock \doarXiv{1712.04454}

\end{thebibliography}
\bibliographystyle{aasjournal}

\end{document}